\documentclass[prl,aps,twocolumn,superscriptaddress,showpacs,floatfix]{revtex4}
\usepackage{amsmath,amssymb,multirow,epsfig,bm,color}

\begin{document} 
\title{Heavy-Light Fermion Mixtures at Unitarity}

\author{Alexandros Gezerlis} 
\affiliation{Theoretical Division, Los Alamos National Laboratory, Los Alamos, New Mexico 87545,  USA}
\affiliation{Department of Physics, University of Illinois at Urbana-Champaign, Urbana, Illinois 61801, USA}
\author{S. Gandolfi} 
\affiliation{S.I.S.S.A., International School of Advanced Studies,
via Beirut 2/4, 34014 Trieste, Italy}
\affiliation{INFN, Sezione di Trieste, Trieste, Italy}
\author{K. E. Schmidt}
\affiliation{Department of Physics, Arizona State University, Tempe, Arizona 85287, USA}
\author{J. Carlson} 
\affiliation{Theoretical Division, Los Alamos National Laboratory, Los Alamos, New Mexico 87545,  USA}

\begin{abstract}
We investigate fermion pairing in the unitary regime for a mass ratio corresponding to a  $^6$Li - $^{40}$K mixture using Quantum Monte
Carlo methods. The ground-state energy and the average light and heavy particle excitation spectrum for the unpolarized superfluid state are nearly independent of the mass ratio. In the majority light
system, the polarized superfluid is close to the energy of a phase separated mixture of nearly fully polarized normal and unpolarized superfluid. For a majority of heavy particles, we find an energy minimum for a normal state with a ratio of $\sim$ 3:1 heavy to light particles. A slight increase in attraction to $k_F a \approx 2.5$ yields a ground state energy of nearly zero for this ratio. A cold unpolarized system in a harmonic trap at unitarity should phase separate into three regions, with a shell of unpolarized superfluid in the middle.
\end{abstract}


\maketitle

Superfluid pairing and the equation of state of cold trapped atoms in the unitary regime have recently been the subject
of intense theoretical and experimental investigation \cite{Giorgini:2008,Ketterle:2008}. These systems are closely related to strongly interacting fermions in
other regimes, such as neutron matter \cite{Gezerlis:2008,Gandolfi:2008} and dense quark matter \cite{Alford:2008},
and hence are useful as prototypes and benchmarks in many areas of physics. 
At unitarity, all physical quantities are simply given by dimensionless numbers times the relevant free Fermi gas
quantity.  Theoretical predictions based upon Quantum Monte Carlo (QMC) calculations
for these dimensionless numbers -- including the ground-state superfluid energy $\xi = E_{sf} / E_{FG} \approx 0.4 $ \cite{Carlson:2003,Astrakharchik:2004,Carlson:2005}, the pairing gap $\eta = \Delta / E_F \approx 0.5$ \cite{Carlson:2003,Carlson:2005,Carlson:2008}, 
and the first-order phase transition between an unpolarized superfluid and a normal
state at finite polarization or concentration $x_c = n_\downarrow/ n_\uparrow \approx 0.44$ \cite{Lobo:2006,Bulgac:2007,Pilati:2008} -- are in good agreement with recent experiments \cite{Luo:2009,Schirotzek:2008,Shin:2008}.

An intriguing variation of this problem is pairing between particles with different 
masses, which is within experimental reach \cite{Wille:2008,Taglieber:2008} and has already 
sparked considerable theoretical interest \cite{Liu:2003,Braaten:2006,Wu:2006,Stecher:2007,Combescot:2007,Baranov:2008,Nishida:2008a,Nishida:2008b,Guo:2008}. 
The most promising candidate is a mixture of $^6$Li and $^{40}$K  s-wave 
Feshbach resonances, for which the mass ratio  $r \approx 6.5$.
A heavy-light fermion mixture may be more likely to exhibit exotic phases, like Larkin-Ovchinnikov-Fulde-Ferrell phases \cite{Bulgac:2008}, 
while for higher mass ratios or more attractive interactions Efimov states are expected to appear.

We consider an interaction of the form:
\begin{equation}
H = \sum_{i=1,N_l}\frac{- \hbar^2}{2 m_l} \nabla_i^2  + \sum_{j=1,N_h} \frac{- \hbar^2}{2 m_h} \nabla_j^2  + \sum_{i,j} V(r_{ij}) ,
\end{equation}
where $h$ denotes a heavy particle and $l$ denotes a light particle, with a mass ratio $ r = m_h / m_l$, and
a zero-range interaction between light and heavy particles with strength tuned to infinite scattering length in the unequal-mass pair.
Mean-field BCS theory for unequal-mass pairing  predicts a simple 
scaling of the equation of state in terms of the reduced mass
$m_r = m_l m_h / (m_l + m_h)$.
If we define the average chemical potential by
$\bar{\mu} = (\mu_{h} + \mu_{l})/2$
then $\bar{\mu}$ and the pairing gap $\Delta$ remain unchanged in units of the reduced Fermi energy
$E_F^{m_r} = \frac{\hbar^2}{4 m_r} (3 \pi^2 n)^{2/3} \equiv \frac{\hbar^2 k_F^2}{4 m_r}~,$ where $n$ is the total particle density.

The heavy and light excitation energies naturally depend upon 
the masses $m_h$ and $m_l$ individually.
The energies of the heavy and light excitations are:
\begin{eqnarray}
E_{h(l)} (k) &= & \frac{ \xi_{h(l)}(k) - \xi_{l(h)}(k)}{2} + \nonumber \\
& & \sqrt{\left( \frac{\xi_{h}(k) 
+ \xi_{l}(k)}{2}\right) ^2 + \Delta^2(k)} , 
\end{eqnarray}
where $\xi_{h(l)}(k) = \frac{\hbar^2 k^2}{2 m_{h(l)}} - \mu_{h(l)}$. 
Even so, the average of $E_{h}(k)$ and $E_{l}(k)$ depends only upon the
reduced mass $m_r$, as does  the gap $\Delta(k)$.

There is no {\it a priori} reason to believe that the BCS results
should be accurate. We have performed QMC 
calculations of the homogeneous superfluid phase, examining the quasi-particle dispersion as a 
function of the momentum. 
The methods are those employed previously in the equal-mass case \cite{Carlson:2005,Carlson:2008}, using
a modified P\"{o}schl-Teller potential with an effective range of $r_0/12$, where $4 / 3 \pi r_0^3 = 1 / n$.
The superfluid and normal phase trial wave functions are of
the same form as used previously, and provide fixed-node upper bounds
to the energy; the superfluid wave function has been variationally re-optimized. 
Of course, new physics corresponding to quite different
nodal strucures of more exotic trial functions (for example LOFF phases)
is not excluded.

For a mass ratio of 6.5, we obtain a ground-state energy $\xi (r=6.5)  = 0.390(5)$, slightly lower
than the $\xi (r=1) = 0.41(1)$ obtained for the same interaction with equal masses.
The latter extrapolates to $\xi (r=1) = 0.40(1)$ at zero effective range; we have
verified that similar small extrapolations are present for unequal masses.
The small difference ($\sim 5 \%$) between the ground-state energies from $r=1$ to $r=6.5$ is a measure of the 
contribution of non-zero total momentum pairs in the ground state.  In QMC calculations
the ground state is determined by a diffusion algorithm which depends upon the local
energy and the mass of the particles.
For a pure BCS state, the local energy $H \Psi_T / \Psi_T$ is identical over the entire $3N$-dimensional coordinate space
for different mass ratios.  However the total mass of the pairs increases with $r$, 
resulting in a slightly lower energy as the diffusion of the pair centers-of-mass
is reduced.

\begin{figure}[t]
\begin{center}
\includegraphics[width=3.5in]{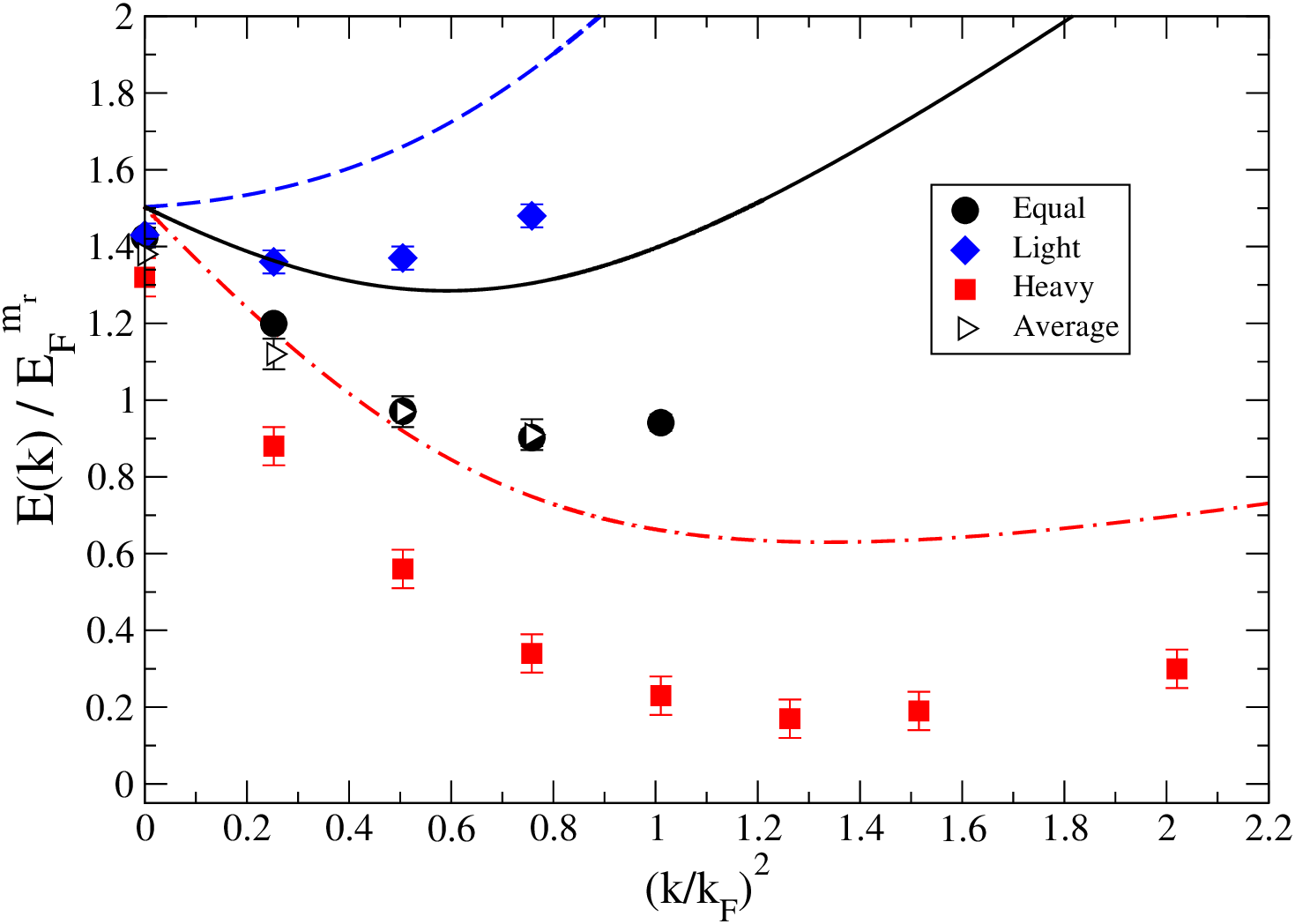}
\caption{(color online) Quasi-particle excitation energies for the two different species. 
The QMC results are shown as diamonds and squares for light and heavy particles, respectively. 
The BCS results are shown as lines: the dashed line corresponds to light particles, 
the dot-dashed line to heavy particles, while the solid line to the results for the equal-mass case. 
Also shown are the QMC results (circles) from Ref. \cite{Carlson:2005} for the 
equal-mass case, as well as the average (triangles) of the two QMC sets of 
points for heavy and light particles.}
\label{fig:particlevsk}
\end{center}
\end{figure}

We have also calculated the quasi-particle excitation energies for the light and heavy particles;
the results are shown in Fig. \ref{fig:particlevsk}. The excitation energies shown for the light (heavy) particles
are calculated by subtracting the energy of $N/2$ heavy and $N/2$ light
particles from the energy using a trial function with
an additional light (heavy) particle in a state of momentum $k$.
Both simulations are performed for the same volume $L^3 = 3\pi^2 N/k_F^3$.
To facilitate comparison with the QMC results, the BCS lines shown are $E_{h}(k) 
+ \mu_{h}$ and $E_{l}(k) + \mu_{l}$ for the heavy and light 
particles, respectively.  Excitation energies for the light particles are higher, and the 
minimum of the quasi-particle dispersion shifts toward zero momentum.

We also compare the average of the light and heavy particle dispersion relations to the
dispersion obtained in the equal-mass case.  The two are very similar, much as they
would be in standard BCS theory.  The superfluid transition temperature would 
nevertheless  decrease with increasing mass ratio, as the average of the minimum 
excitation energies in each branch is significantly lower than the minimum energy
at equal masses.  We find this average (calculated by subtracting $\bar{\mu}$) to be $\eta (r=6.5) = 0.38(4)$, in comparison
to the calculated value $\eta (r=1) =0.50(5)$ in the equal-mass case.  The individual
spectra are also important. For example, radio frequency response experiments, which have been used to
explore the gap in the equal-mass case \cite{Schirotzek:2008}, could be designed to be sensitive to the individual light and heavy
particle dispersion relations.

\begin{figure}[t]
\begin{center}
\includegraphics[width=3.5in]{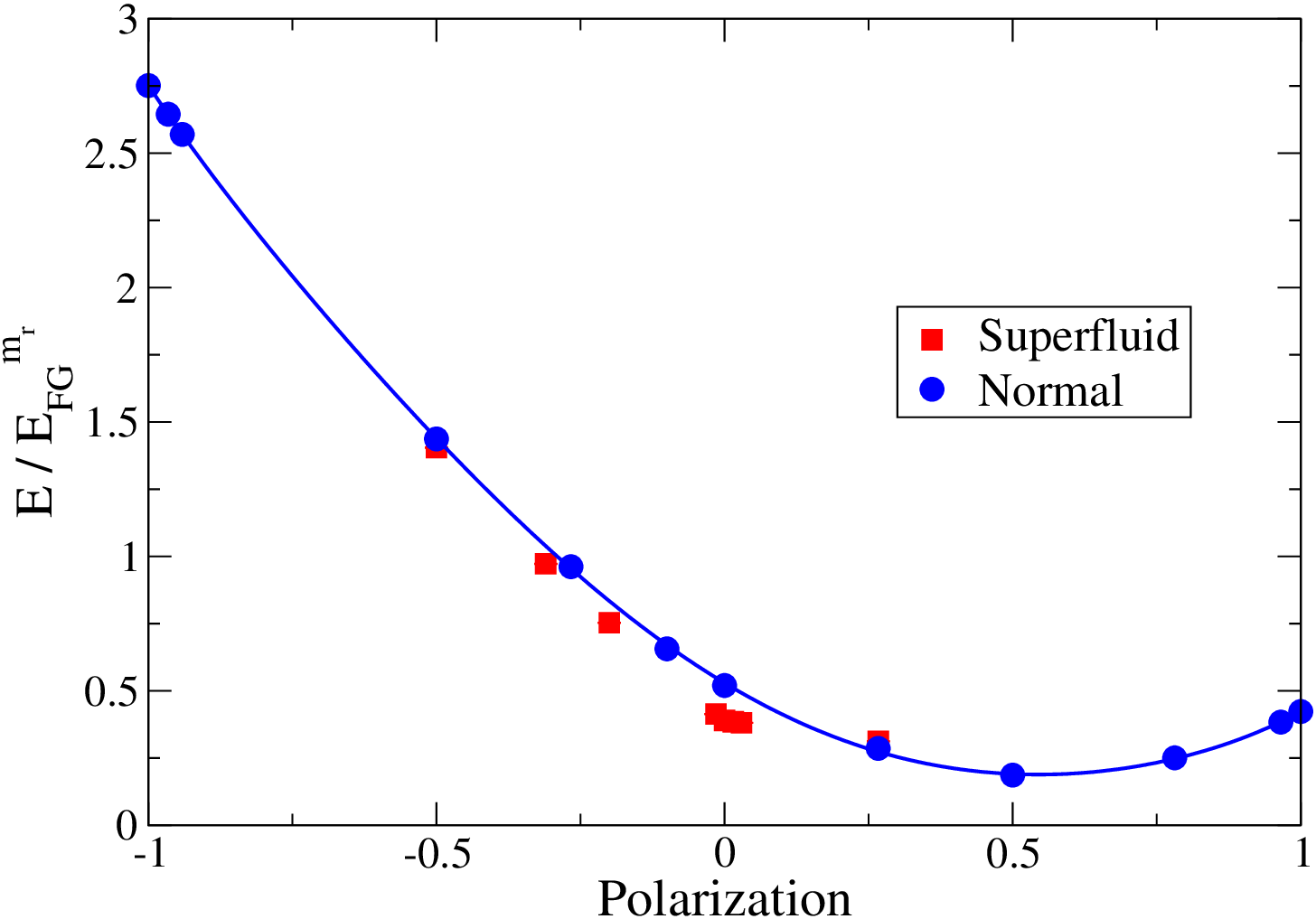}
\caption{(color online) Equation of state of a heavy-light fermion mixture as a function 
of the polarization at constant density. Shown are results for the normal state (circles) as well as for the 
superfluid state (squares). 
The solid line is a polynomial best fit to the QMC results for the normal state.}
\label{fig:vspol}
\end{center}
\end{figure}

Away from equal populations, we explore the phase diagram at zero temperature
by considering normal and gapless superfluid states using from 60 to 90 particles.
For calculations of the normal state, the trial wave functions dictating the nodes
are taken from free-particle Slater determinants with filled-shell configurations
in periodic boundary conditions. In Fig. \ref{fig:vspol} we plot the ground-state energy versus 
the polarization $P = (N_h - N_l)/ (N_h + N_l)$, in units of $E_{FG}^{m_r}$. 

The circles are QMC calculations of the normal state and the curve is a simple polynomial fit to the normal
state results as a function of polarization.  The polynomial fit is explicitly tied to the
free-particle results at $P=\pm 1$, and to the binding energy $B_{h(l)}$ of a single heavy (light)
particle in a sea of light (heavy) particles. With the majority particle number $N_{l(h)}$ and the simulation volume $L^3$ constant,
we find $B_h = 0.36 E_F^l  = 0.99(5) E_F^{m_r} $ and  $B_l = 2.3 E_F^h = 0.97(5) E_F^{m_r}$ at $r=6.5$.
With this definition the equal mass binding $B = 0.6 E_F$.
At constant total densities, these results correspond to $B_h = 0.76 E_F^l$ and $B_l = 2.7 E_F^h $,
in rough agreement with results in Ref. \cite{Combescot:2007}.
By fitting the dispersion of single impurities, we find $m_l^\star / m_l = 1.3$ and $m_h^\star / m_h = 1.0$.

We find an energy minimum near a polarization
of 0.5, corresponding to a ratio of 3:1 heavy to light particles. The small value of the
energy  indicates a possible collapse of the normal state at a mass ratio 
smaller than that found in 3-body calculations, where Efimov states and a collapse
begin at a mass ratio of $13.6$ \cite{Braaten:2006,Nishida:2008a}.  At unitarity with this finite range
potential we observe collapse (large negative energies and several particles within the interaction range) before $r=12$.  At a mass ratio $r=6.5$, we find that the
energy decreases quickly with interaction strength, reaching zero at $k_F a \approx 2.5$.
These few-particle correlations may increase loss rates and limit the effectiveness of standard cooling techniques which sweep from the Bose-Einstein condensation regime to the unitary regime.
As the mass ratio increases, the minimum in energy
will shift toward higher polarizations.  It would be very interesting to examine this evolution
and the associated Fermi or Bose condensates of odd and even clusters of fermions.

We also consider the possibility of polarized superfluids; a simple case is the gapless
superfluid where unpaired particles are placed  at the minimum
of the dispersion curves in Fig. \ref{fig:particlevsk}. The energies for the gapless superfluid
state are shown as squares in Fig. \ref{fig:vspol}. Over a range of polarizations $P<0$,
we find the polarized superfluid has a significantly lower energy than the homogeneous normal state at the
same density and polarization.

\begin{figure}[t]
\begin{center}
\includegraphics[width=3.5in]{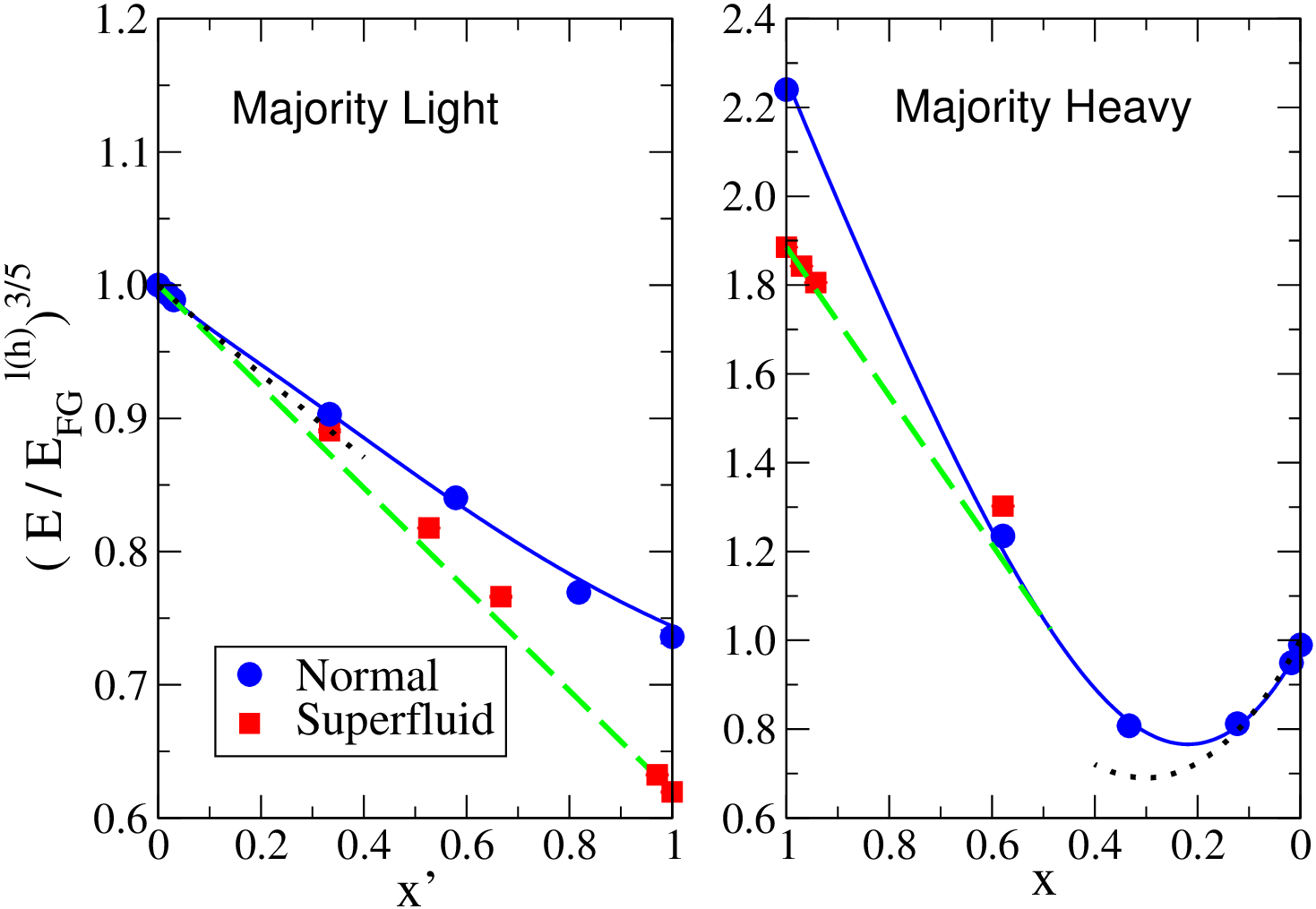}
\caption{(color online) Equation of state 
as a function of the concentration $x'$ (left panel) and  $x$ (right panel). 
Normal (circles) and superfluid states (squares) are shown.
The dashed lines are the coexistence lines between the normal and the unpolarized superfluid states. 
The resulting critical concentrations are $x'_c = 0.02$ and $x_c = 0.49$.
The dotted curves represent non-interacting impurities
with the calculated binding energies and effective masses.}
\label{fig:vsx}
\end{center}
\end{figure}

The results displayed in Fig. \ref{fig:vspol} can be used to determine the stability of these phases.
We use a polynomial fit to the normal state to calculate the critical
concentrations of heavy and light particles and possible
first-order phase transitions that occur between the
superfluid and the normal states at finite polarization. This is illustrated in
Fig. \ref{fig:vsx}, where we plot the energy normalized to the free-particle energy of
the majority species at the relevant density to the 3/5 power as a function of the concentration $x' = n_h/n_l$ for the majority light particle
case and $x = n_l/n_h$ for the majority heavy particle case.
The transition points can be found by equating the pressures and chemical potentials
of the normal and superfluid states.  For the case of majority light species, the
equilibrium concentration of heavy particles is extremely small, $x'_c = 0.02(2)$,
indicating equilibrium between a superfluid and a nearly fully polarized sea of light particles.
For the majority heavy case the critical concentration is $x_c = 0.49(5)$, 
near the concentration found for the equal mass case: $x_c = 0.44$ \cite{Lobo:2006}.
The transitions indicated in the figure are calculated from the polynomial fits and indicated
by dashed lines following the tangent construction used in Ref. \cite{Bulgac:2007}.

The polarized superfluid results are also shown in Fig. \ref{fig:vsx}.  Over a range of polarizations
$P < 0$ these states are very close to stability with respect to the phase separated
normal state and unpolarized superfluid.  It is possible that further generalizations
of the trial states, for example by considering inhomogeneous polarized superfluids like
LOFF states, would lower the energy  and provide a stable polarized superfluid at
zero temperature.

The pressures and chemical potentials calculated for the superfluid and normal states
can also be used to examine what happens in a harmonic trap.  Keeping the chemical
potentials $\mu^0_{h(l)}$ fixed  and choosing the state of highest pressure 
with  local chemical potentials $\mu_{h(l)} (r) = \mu ^0_{h(l)}  - V_{h(l)} (r)$ 
one can calculate, within the local-density approximation, the density for each species in the trap. 
In general, the
trapping potentials of the two species are unequal; for this analysis we assume 
harmonic potentials with a strength  $ m_h \omega_h^2 / 2 $ for the heavy
particles equal to twice that of the light potential strength $m_l \omega_l^2/2$,
similar to that of a recent experiment on $^6$Li - $^{40}$K mixtures \cite{Wille:2008}.

\begin{figure}[t]
\begin{center}
\includegraphics[width=3.5in]{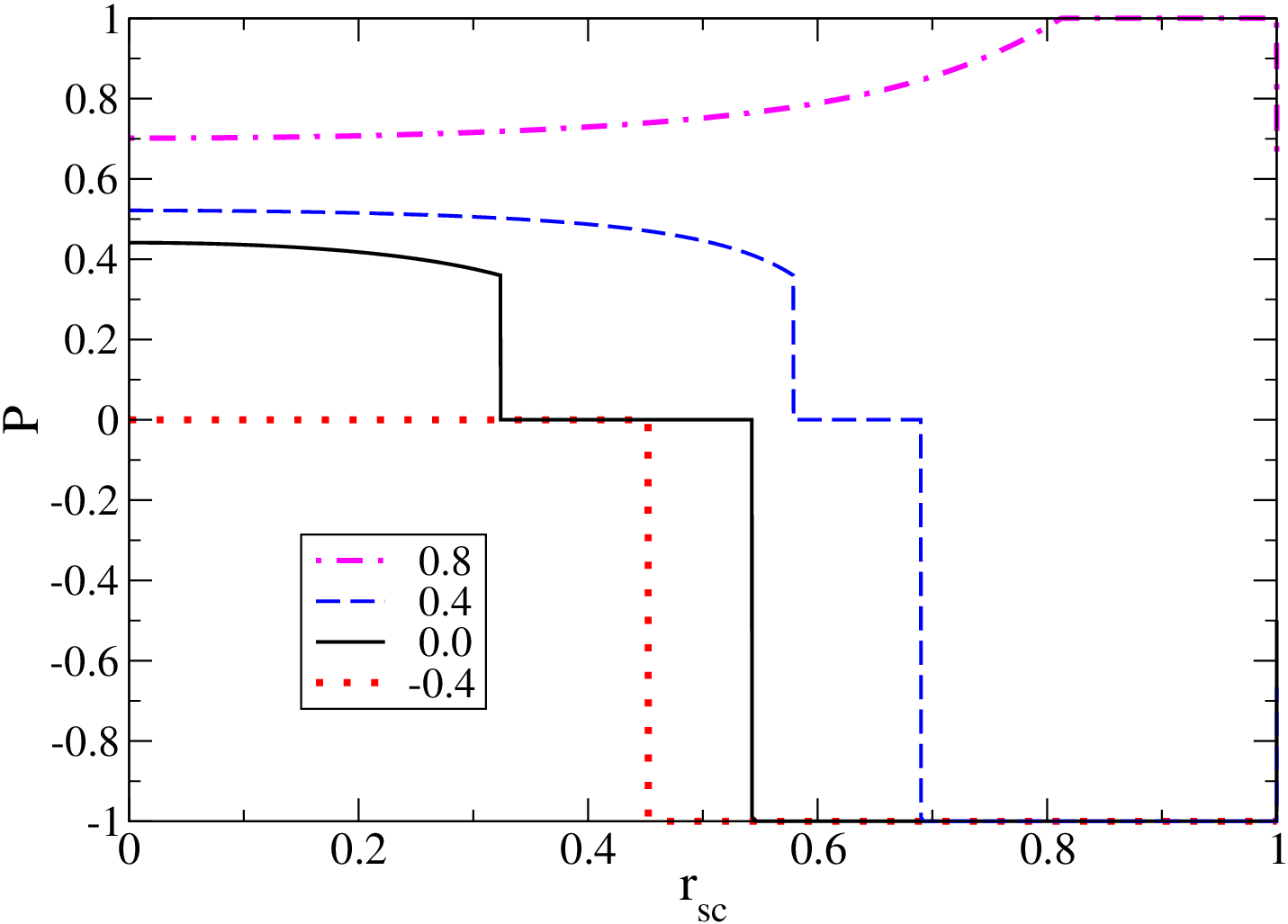}
\caption{(color online) The polarization of a trapped system  as a function of the  radius (scaled so that at $r_{sc}=1$ the density goes to 0). Shown are curves for four different values of $P_{tot} = (N_{h} - N_{l})/(N_{h} + N_{l}).$}
\label{fig:polarization}
\end{center}
\end{figure}

In Fig. \ref{fig:polarization} we plot the local polarization as a function of scaled radius
for various total polarizations $ P_{tot} = (N_h - N_l)/ (N_h+N_l)$ where $N_h$ and $N_l$ are the
total number of heavy and light particles in the trap.  Curves are shown for $P_{tot} = $ -0.4, 0, 0.4, and 0.8.  
The radii are scaled in each case so that within the local density approximation the density falls to zero at $r_{sc}=1$.
In this plot we assume that  the polarized superfluid is unstable at $T=0$.
For large total negative polarizations, the equilibrium configuration is
an unpolarized ($P=0$) superfluid in the center and a nearly fully polarized sea of light particles
in the exterior.  At a finite temperature the polarized superfluid state at $P<0$ would appear,
similar to what happens in the equal mass case. Because of the small energy differences,
we expect finite temperature effects to be even more important here.

Near zero total polarization, in contrast, three distinct regions exist. In the center a normal
state is favored with a polarization near $P=0.5$; this is the lowest-energy normal state of the system
described earlier.  At larger radius, there is a shell of unpolarized superfluid, and then again
in the exterior a region of nearly all light particles, or potentially a polarized superfluid.  At this large mass ratio, 
this unusual configuration is actually lower in energy than a homogeneous superfluid everywhere.
Finally, for very large total polarizations the system is normal everywhere, with the polarization
smoothly increasing from the center of the trap as the radius increases.  This is again analogous
to what happens in the equal-mass case for a large total polarization.
It would be interesting to confirm this new structure experimentally, to explore the stability
and structure of the polarized superfluid, and to determine
its evolution with mass ratio.

In summary, we performed  QMC studies of heavy-light fermion mixtures at unitarity. 
We find that the ground-state energy of the superfluid
and the average quasi-particle dispersion
agrees closely with the superfluid with equal masses and the same reduced mass $m_r$,
as predicted by BCS theory.  In contrast, 
the system at finite polarization is very different from the equal-mass case, resulting in
significantly different profiles of trapped systems, even for the case of equal numbers
of heavy and light particles.  Polarized superfluids are very near to stability for more
light than heavy particles. In the majority heavy case, we find an energy minimum
at approximately a 3:1 heavy to light ratio which evolves rapidly with mass ratio and
interaction strength.
The extra scale made available by the mass ratio produces a variety
of fascinating new physical effects in cold Fermi atoms near unitarity.

The authors wish to thank A. Trombettoni, S. Giorgini, S. Pilati, S. Reddy, D. Son and H. T. C. Stoof
for stimulating discussions. 
The work of A.G. and J.C. is supported by the Nuclear Physics Office of the U.S. Department 
of Energy under Contract No. DE-AC52-06NA25396 and by the LDRD program at Los Alamos National Laboratory.  
Computing resources were 
provided at LANL through the Institutional Computing Program and at NERSC. 
Calculations by S.G. were partially performed on the BEN cluster at ECT* in Trento, 
under a grant for supercomputing projects, and partially on the HPC facility of 
SISSA/Democritos in Trieste. 
The work of K.E.S. was supported in part by NSF Grant No. PHY07-57703.
The work of A.G. was supported in part by NSF Grant Nos. PHY03-55014 and PHY07-01611. 

{\it Note added in proof}.-- An analysis of the phase diagram of Fermi mixtures at unitarity was recently published \cite{Bausmerth:2009}. 
It would be interesting to see this calculation repeated for the equation of state that we have obtained.


\begin{thebibliography}{99}

%Theory of ultracold atomic Fermi gases
\bibitem{Giorgini:2008} S. Giorgini, L. P. Pitaevskii, and S. Stringari, Rev. Mod. Phys. {\bf 80}, 1215 (2008).

%Making, probing and understanding ultracold Fermi gases
\bibitem{Ketterle:2008} W. Ketterle and M. W. Zwierlein, in {\it Ultracold Fermi Gases}, Proceedings of the International School of Physics ``Enrico Fermi,'' Course CLXIV, edited by M. Inguscio, W. Ketterle, and C. Salomon (IOS Press, Amsterdam, 2008). [arXiv:0801.2500]

%Strongly paired fermions: Cold atoms and neutron matter
\bibitem{Gezerlis:2008} A. Gezerlis and J. Carlson, Phys. Rev. C, {\bf 77} 032801 (2008).

% Equation of state of superfluid neutron matter and the calculation of S(0)-1 pairing gap.
\bibitem{Gandolfi:2008} S. Gandolfi {\it et al.}, Phys. Rev. Lett. {\bf 101}, 132501 (2008)

%Color superconductivity in dense quark matter
\bibitem{Alford:2008} M. G. Alford {\it et al.}, Rev. Mod. Phys. {\bf 80}, 1455 (2008).

%Superfluid Fermi Gases with Large Scattering Length
\bibitem{Carlson:2003} J. Carlson {\it et al.}, Phys. Rev. Lett. {\bf 91}, 050401 (2003).

%Equation of State of a Fermi Gas in the BEC-BCS Crossover: A Quantum Monte Carlo Study
\bibitem{Astrakharchik:2004} G. E. Astrakharchik {\it et al.}, Phys. Rev. Lett. {\bf 93}, 200404 (2004).

%Asymmetric Two-Component Fermion Systems in Strong Coupling
\bibitem{Carlson:2005} J. Carlson and S. Reddy, Phys. Rev. Lett. {\bf 95}, 060401 (2005).

%Superfluid Pairing Gap in Strong Coupling
\bibitem{Carlson:2008} J. Carlson and S. Reddy, Phys. Rev. Lett. {\bf 100}, 150403 (2008).

%Normal State of a Polarized Fermi Gas at Unitarity
\bibitem{Lobo:2006} C. Lobo {\it et al.}, Phys. Rev. Lett. {\bf 97}, 200403 (2006).

%Zero-temperature thermodynamics of asymmetric Fermi gases at unitarity
\bibitem{Bulgac:2007} A. Bulgac and M. M. Forbes, Phys. Rev. A {\bf 75}, 031605 (2007).

%Phase Separation in a Polarized Fermi Gas at Zero Temperature
\bibitem{Pilati:2008} S. Pilati and S. Giorgini, Phys. Rev. Lett. {\bf 100}, 030401 (2008).

%Thermodynamic Measurements in a Strongly Interacting Fermi Gas
\bibitem{Luo:2009} L. Luo and J. E. Thomas, J. Low Temp. Phys. {\bf 154}, 1 (2009).

%Determination of the Superfluid Gap in Atomic Fermi Gases by Quasiparticle Spectroscopy
\bibitem{Schirotzek:2008} A. Schirotzek {\it et al.}, Phys. Rev. Lett. {\bf 101}, 140403 (2008).

%Phase diagram of a two-component Fermi gas with resonant interactions
\bibitem{Shin:2008} Y. Shin {\it et al.}, Nature {\bf 451}, 689 (2008).

%Quantum Degenerate Two-Species Fermi-Fermi Mixture Coexisting with a Bose-Einstein Condensate
\bibitem{Taglieber:2008} M. Taglieber {\it et al.}, Phys. Rev. Lett. {\bf 100}, 010401 (2008).

%Exploring an Ultracold Fermi-Fermi Mixture: Interspecies Feshbach Resonances and Scattering Properties of 6Li and 40K
\bibitem{Wille:2008}  E. Wille {\it et al.}, Phys. Rev. Lett. {\bf 100}, 053201 (2008).

%Interior Gap Superfluidity
\bibitem{Liu:2003} W. V. Liu and F. Wilczek, Phys. Rev. Lett. {\bf 90} 047002 (2003).

%Universality in few-body systems with large scattering length
\bibitem{Braaten:2006} E. Braaten and H.-W. Hammer, Phys. Rept. {\bf 428} 259 (2006).

%Resonant pairing between fermions with unequal masses
\bibitem{Wu:2006} S.T. Wu, C.-H. Pao, and S.-K. Yip, Phys. Rev. B {\bf 74}, 224504 (2006)

%BEC-BCS crossover of a trapped two-component Fermi gas with unequal masses
\bibitem{Stecher:2007} J. von Stecher, C. H. Greene, and D. Blume, Phys. Rev. A {\bf 76}, 053613 (2007).

%Normal State of Highly Polarized Fermi Gases: Simple Many-Body Approaches
\bibitem{Combescot:2007} R. Combescot {\it et al.}, Phys. Rev. Lett. {\bf 98}, 180402 (2007).

%Superfluid pairing between fermions with unequal masses
\bibitem{Baranov:2008} M. A. Baranov, C. Lobo, and G. V. Shlyapnikov, Phys. Rev. A {\bf 78}, 033620 (2008).

%Universal Fermi Gas with Two- and Three-Body Resonances
\bibitem{Nishida:2008a} Y. Nishida, D. T. Son, and S. Tan, Phys. Rev. Lett. {\bf 100}, 090405 (2008).

%Casimir interaction among heavy fermions in the BCS-BEC crossover
\bibitem{Nishida:2008b} Y. Nishida, Phys. Rev. A {\bf 79}, 013629 (2009).

%Finite-Temperature Behavior of an Inter-species Fermionic Superfluid with Population Imbalance
\bibitem{Guo:2008} H. Guo {\it et al.}, Phys. Rev. A 80, 011601(R) (2009).

%Unitary Fermi Supersolid: The Larkin-Ovchinnikov Phase
\bibitem{Bulgac:2008} A. Bulgac and M. M. Forbes, Phys. Rev. Lett. {\bf 101}, 215301 (2008).

%Chandrasekhar-Clogston limit and phase separation in Fermi mixtures at unitarity
\bibitem{Bausmerth:2009} I. Bausmerth, A. Recati, and S. Stringari, Phys. Rev. A {\bf 79}, 043622 (2009).

\end{thebibliography}
\end{document}